\documentclass{article}
\usepackage{spconf,amsmath,graphicx,amssymb,xcolor,hyperref,enumitem}
\usepackage[flushleft]{threeparttable}
\usepackage{array,mathtools}
\usepackage{multirow,setspace}
\newcolumntype{P}[1]{>{\centering\arraybackslash}p{#1}}

\title{OSSEM: one-shot speaker adaptive speech enhancement \\ using meta learning}
\name{Cheng Yu$^{1}$, Szu-Wei Fu$^{1}$, Tsun-An Hsieh$^{1}$, Yu Tsao$^{1}$, and Mirco Ravanelli$^{2}$
}
\address{$^{1}$Research Center for Information Technology Innovation, Academia Sinica, Taiwan \\
$^{2}$Mila-Quebec AI Institute, Montreal, Canada\\}
\begin{document}
\ninept
\maketitle
\begin{abstract}
Although deep learning (DL) has achieved notable progress in speech enhancement (SE), further research is still required for a DL-based SE system to adapt effectively and efficiently to particular speakers. In this study, we propose a novel meta-learning-based speaker-adaptive SE approach (called OSSEM) that aims to achieve SE model adaptation in a one-shot manner. OSSEM consists of a modified transformer SE network and a speaker-specific masking (SSM) network. In practice, the SSM network takes an enrolled speaker embedding extracted using ECAPA-TDNN to adjust the input noisy feature through masking. To evaluate OSSEM, we designed a modified Voice Bank-DEMAND dataset, in which one utterance from the testing set was used for model adaptation, and the remaining utterances were used for testing the performance. Moreover, we set restrictions allowing the enhancement process to be conducted in real time, and thus designed OSSEM to be a causal SE system. Experimental results first show that OSSEM can effectively adapt a pretrained SE model to a particular speaker with only one utterance, thus yielding improved SE results. Meanwhile, OSSEM exhibits a competitive performance compared to state-of-the-art causal SE systems. 

\end{abstract}
\begin{keywords}
Speech Enhancement, Speaker Embedding, Meta-learning, Deep Learning 
\end{keywords}
\section{Introduction}
\label{sec:intro}

The goal of speech enhancement (SE) is to improve the quality and intelligibility of distorted speech. In various speech-related applications, such as automatic speech recognition \cite{donahue2018exploring}, speaker recognition \cite{Sper1,SPer2}, and assistive hearing devices \cite{puder2009hearing}, SE serves as an indispensable front-end unit. Traditional SE methods \cite{priori2,priori1} are derived based on statistical models and assumed properties of speech and noise signals. Under scenarios in which the assumptions are unsatisfactory, the traditional SE approaches may yield a suboptimal performance.
\par
Deep learning (DL) models have recently been widely applied for SE tasks \cite{DDAE,FCN, SEDNN, LSTM1}. 
Although several studies have shown that DL-based SE systems can outperform traditional methods, their limited generalization ability is still an issue. One simple solution to improve the generalization of the model involves collecting as much training data as possible. However, it is almost impossible to cover all types of conditions (e.g., different SNRs, noise types, and speakers). Another solution is to adapt the SE model before/during the inference using auxiliary information. 
For example, in \cite{yu2020speech, kim2021test, zezario2020speech}, the authors first prepared multiple SE models, and then trained them under different conditions. In the testing stage, a most-match model was selected based on certain criteria. 
\par
Another class utilizes additional noise or speaker information to guide the SE adaptation to certain types of noise \cite{xu2014dynamic, liao2018noise, li2020noise, lee2020dynamic} or speaker \cite{chuang2019speaker} conditions. In this study, we go one step further to simultaneously adapt both the model inputs and model weights through embedding vector and meta-learning, respectively. Specifically, we propose a novel one-shot speaker-adaptive SE approach using meta-learning (OSSEM), in which the SE model can be effectively and efficiently adapted to a particular speaker with only one adaptation sample. OSSEM consists of two networks: a modified transformer SE network that aims to achieve an SE and a speaker-specific masking (SSM) network that generates speaker-specific masks for adapting the SE model. For the SSM network, we adopted ECAPA-TDNN \cite{desplanques2020ecapa} through the SpeechBrain toolkit \cite{speechbrain} to extract speaker embeddings as the input features. The two networks were trained in a meta-learning manner, which has been shown to be effective for one/few-shot learning on several tasks \cite{hsu2020meta, shi2020few}. In contrast to previous studies that adapt the overall SE model \cite{zhou2021meta}, OSSEM can reach a fast adaptation because only the parameters in the SSM network are adjusted instead of the entire OSSEM system. In addition, we propose several training techniques for OSSEM to further improve its stability and performance.
\par
To evaluate the proposed OSSEM system, we slightly modified the Voice Bank-DEMAND \cite{vctkb} dataset. One utterance from each speaker was 
selected from the testing set for model adaptation, and the remaining utterances from the same speaker were used for testing the performance. Experiment results confirmed the fast model adaptation of OSSEM and showed that it can yield a competitive performance comparable to that of state-of-art casual SE systems. The remainder of this paper is organized as follows.
Section \ref{sec:related} reviews related studies and previous meta-learning. Section \ref{sec:method} then elaborates on the proposed OSSEM approach. Section \ref{sec:ERA} reports the experiment setup and results. Finally, Section \ref{sec:conclude} provides some concluding remarks regarding the findings and contributions of this study.

\vspace{-0.3cm}

\begin{figure*}[t]
    \centering
    \includegraphics[width=16cm]{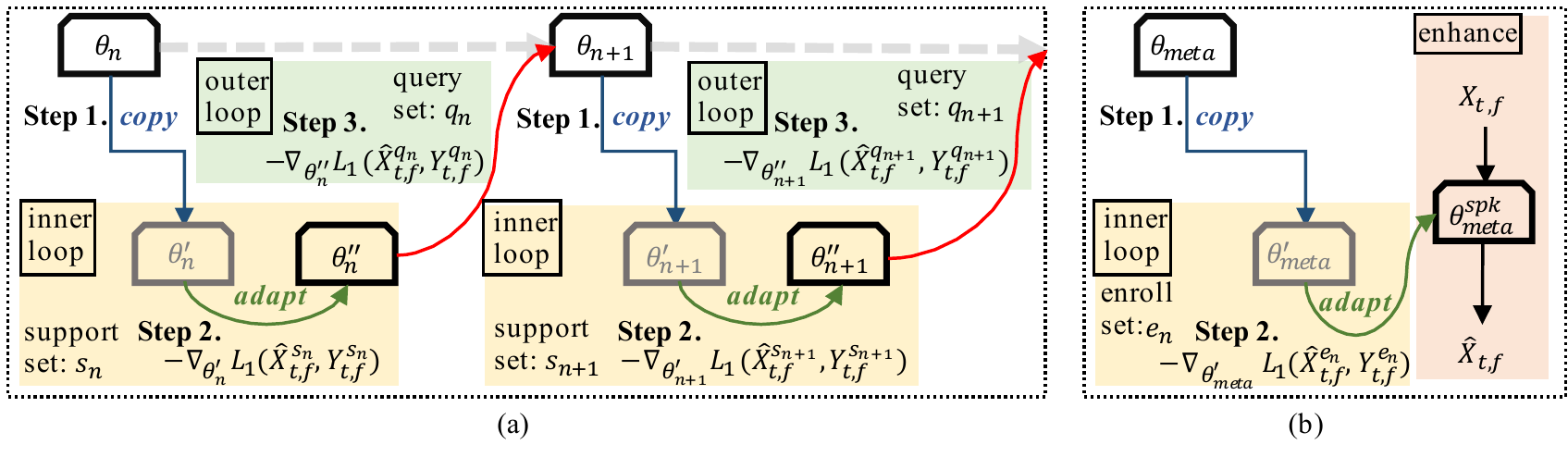}
    \vspace{-0.5cm}
    \caption{Flow chart of the OSSEM system, (a) the training stage (offline), and (b) the one-shot learning and testing stages (online).}
    \label{fig:Training}
\end{figure*}

\section{Meta-learning SE system}
\label{sec:related}

The aim of OSSEM is to effectively adapt a pretrained SE model to a particular speaker in a one-shot learning manner. To this end, as the main criterion for training OSSEM, we adopt the meta-learning algorithm, which has been popularly used to build one/few-shot machine learning systems \cite{finn2017model}. 
To apply meta-learning to SE systems, the training set data needs to be reorganized into task-specific classes, such as noise conditions \cite{zhou2021meta} or speaker identities (considered in this study).
Each class is further divided into support and query set. The former is used for adaptation by one (or few) shot learning, whereas the latter is used for model training. To the best of our knowledge, the proposed OSSEM is the first one-shot speaker-adaptation SE using meta-learning. To clarify the implementation details, we elaborate on the training, one-shot learning, and testing stages as follows. 
\subsection{Training stage (offline)}
\label{ssec:training stage}
Fig. \ref{fig:Training} (a) demonstrates the training stage (offline) of OSSEM. For clarity, we denote the DL model by $g_{\theta_n}$, with weights $\theta_n$ of the $n$-th iteration. We elaborate three steps in this stage as follows: In \textbf{Step 1}, the weights $\theta_{n}$ are copied as $\theta^{'}_{n}$. In \textbf{Step 2}, the weights $\theta^{'}_{n}$ are adapted using gradients from the training on the support set $s_{n}$, where $\hat{X}^{s_{n}}_{t,f} = g_{\theta^{'}_{n}}(X^{s_{n}}_{t,f})$ denotes the prediction using the input feature $X^{s_{n}}_{t,f}$, and $Y^{s_{n}}_{t,f}$ denotes the clean features. Finally, in \textbf{Step 3}, the adapted weights $\theta^{''}_{n}$ are then trained using the corresponding query set $q_{n}$, where $\hat{X}^{q_{n}}_{t,f} = g_{\theta^{''}_{n}}(X^{q_{n}}_{t,f})$ denotes the prediction using the input feature $X^{q_{n}}_{t,f}$, and $Y^{q_{n}}_{t,f}$ denotes the clean features. The gradients computed from this step are used to update the original weights $\theta_{n}$ to yield $\theta_{n+1}$. The two gradients in \textbf{Step 2} and \textbf{Step 3} are computed using the losses as follows:
\vspace{-0.1cm}
\begin{gather}
    L_1(\hat{X}^{s_{n}}_{t,f}, Y^{s_{n}}_{t,f}) = \frac{1}{N(s_n)}\sum^{N(s_{n})}_{n=1}\|\hat{X}^{s_{n}}_{t,f}-Y^{s_{n}}_{t,f}\|_{1} \\
    L_1(\hat{X}^{q_{n}}_{t,f}, Y^{q_{n}}_{t,f}) = \frac{1}{N(q_n)}\sum^{N(q_{n})}_{n=1}\|\hat{X}^{q_{n}}_{t,f}-Y^{q_{n}}_{t,f}\|_{1}
\end{gather}
where $N(s_{n})$ and $N(q_{n})$ denote the numbers of data in the support and query sets, respectively. The value $N(s_{n})$ is usually extremely small (e.g., in this paper, $N(s_{n})$ = 1 for one-shot learning and $N(q_{n})$ = 20 for \textbf{Step 3}). The inner and outer loops represent supervised training cycles on the support set $s_{n}$ and query set $q_{n}$, respectively.

\subsection{One-shot learning and testing stage (online)}
\label{ssec:testing stage}

We demonstrate the testing stage (online) flowchart of the OSSEM in Fig. \ref{fig:Training} (b). We elaborate on two steps in this stage as follows: In \textbf{Step 1}, the weights $\theta_{meta}$ are copied as $\theta^{'}_{meta}$. This copy step allows users to maintain well-trained weights $\theta_{meta}$, whereas adaptation is only applied on the copied weights. In \textbf{Step 2}, the weights $\theta^{'}_{meta}$ are adapted using gradients from the training on the enrollment set $e_{n}$, where $\hat{X}^{e_{n}}_{t,f} = g_{\theta^{'}_{meta}}(X^{e_{n}}_{t,f})$ and $Y^{e_{n}}_{t,f}$ denotes the clean features. Finally, the adapted weights $\theta^{spk}_{meta}$ are ready for SE to enhance the noisy features $X_{t,f}$ to $\hat{X}_{t,f}$. The gradients in \textbf{Step 2} are computed using the loss as follows:
\vspace{-0.1cm}
\begin{gather}
    L_1(\hat{X}^{e_{n}}_{t,f}, Y^{e_{n}}_{t,f}) = \frac{1}{N(e_n)}\sum^{N(e_{n})}_{n=1}\|\hat{X}^{e_{n}}_{t,f}-Y^{e_{n}}_{t,f}\|_{1}
\end{gather}
where $N(e_{n})$ denotes the number of data in the enrollment set, which is usually an extremely small number. For example, for a one-shot adaptation, $N(e_{n})$ = 1. In the proposed OSSEM system, the gradients from Eq. (3) achieve an efficient adaptation because $\nabla_{\theta^{'}_{meta}}L_1(\hat{X}^{e_{n}}_{t,f}, Y^{e_{n}}_{t,f})$ only adapts to the SSM network of OSSEM.






\vspace{-0.3cm}
\begin{figure}[h]
    \centering
    \includegraphics[width=6.0cm]{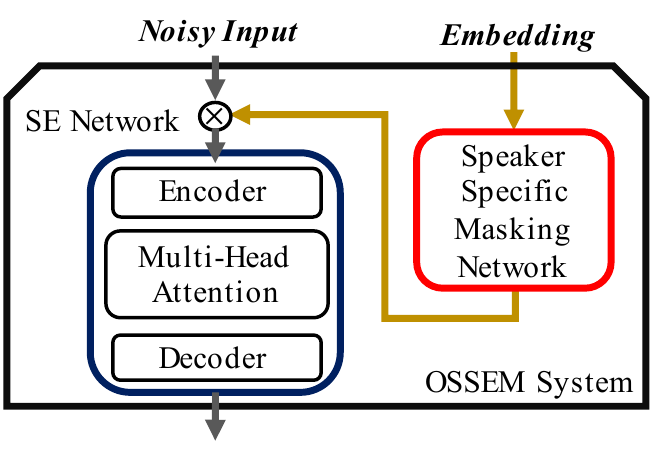}
    \vspace{-0.5cm}
    \caption{Model architecture of the OSSEM system that consists of a modified Transformer (encoder, multi-head attention, and decoder) SE network and an SSM network.}
    \label{fig:BK diagram}
    \vspace{-0.5cm}
\end{figure}

\section{Proposed OSSEM system}
\label{sec:method}

\subsection{Model architecture}
\label{ssec:semodule}
\subsubsection{Modified Transformer-based SE network}
\label{sssec:modifiedtransformer}
The Transformer model \cite{vaswani2017attention} has recently been used in SE systems and has shown a promising performance \cite{kim2020transformer, fu2020boosting}.
In the proposed OSSEM system, we adopted the modified transformer model proposed in \cite{fu2020boosting}; the model consists of a convolutional encoder (for replacing the original positional encoder), several multi-head attention blocks, and a fully connected layer. The convolutional encoder includes four 1-D convolutional layers, and each attention block consists of eight heads, with 64 dimensions per head.

\vspace{-0.3cm}
\subsubsection{The SSM Network}
\label{sssec:SMN}
As shown in Fig. \ref{fig:BK diagram}, SSM takes the speaker embedding (from ECAPA) as input and generates speaker-specific masks. The SSM network consists of a three-layered dense network with the leaky-Relu activation functions for the first two layers and the sigmoid function for the last layer. The generated mask is then multiplied with the noisy spectrogram to yield speaker-adaptive features, which are the inputs to the Transformer-based SE network. 

\subsection{Training techniques}
\label{ssec:OSSEMmethod}
In our preliminary experiments, directly applying meta-learning to SE tasks did not achieve good results, probably because a model trained using MAML becomes easily unstable \cite{antoniou2018train}.). To optimally exert the capability of the proposed OSSEM system, we propose several training techniques and describe them as follows:

\begin{enumerate}[leftmargin=\parindent,align=left,labelwidth=\parindent,labelsep=0pt]
    \item \textbf{Speaker-Inner-Loop:} \\
    MAML was previously reported to suffer from a trade-off \cite{antoniou2018train} between burdensome inner loops and better performance. To solve this problem, Raghu et al. proposed an almost-no-inner-loop (ANIL) \cite{raghu2019rapid} that only updates task-specific parameters in the inner loop. However, it may be difficult to determine the task-specificity of the parameters in a model without a predefined task-specific network. In this study, the OSSEM system consists of two task-specific building blocks. The modified transformer SE network is responsible for general-purpose SE, whereas the SSM network adapts the input features based on a specific speaker. We propose a speaker-inner-loop method to apply adaptation only on the SSM network in the inner loop. As shown in Fig. \ref{fig:Training} (a), the gradients   $\nabla_{\theta^{'}_{n}}L_{1}(\hat{X}^{s_{n}}_{t,f}, Y^{s_{n}}_{t,f})$ (computed using support set data $s_{n}$ of a specific speaker) are only used to adapt the parameters of the SSM network and keep the parameters in the SE network unchanged. The speaker-inner-loop is beneficial for two reasons: First, the updates are completely focused on the adaptation of the task-specific SSM network parameters. Second, the adaptation efficiency can be further improved by focusing only on the compact SSM network instead of the entire OSSEM system.
    
    \item \textbf{Learning Rate Scaling Rule:} \\
    As shown in Fig. \ref{fig:Training} (a), the training stage of OSSEM has an inner-loop and an outer-loop. Each loop has a specific learning rate (i.e., inner-loop learning rate (ILR) and outer-loop learning rate (OLR)). For OLR, we fixed its value. For ILR, we adopted a modified learning rate scaling rule \cite{goyal2017accurate} to stabilize the training procedure. Note that when ILR is $0$ (removing \textbf{Step 2} in Fig. \ref{fig:Training} (a)), the training stage of OSSEM becomes a normal supervised training. For each outer loop, the training data in the query set are from the same speaker, which may make model training difficult. Thus, as in \cite{liao2018noise}, the weights of a normally supervised trained model are applied as the weight initialization for OSSEM. To smoothly transform from supervised training to meta-learning, we set ILR to $0$ in the first five epochs. From the sixth epoch, ILR was linearly scaled up (warmup) and then fixed during the last 20 epochs to stabilize the training.

    \item \textbf{Feature Re-scaling Inner-Loop:} \\
    Feature re-scaling has been confirmed to be effective \cite{grus2019data} at stabilizing the gradients when training DL-based tasks. Because OSSEM relies on the effectiveness of speaker adaptation, it is crucial to stabilize the training of the inner loop. Specifically, speaker adaptation functions through an SSM network, which is highly dependent on the training of the inner loop. To further stabilize the training of the inner loop, we designed a feature re-scaling ratio $\alpha$ that is computed between clean and noisy features (the average energy power ratio between features):
    \begin{gather}
        \alpha = \frac{\|Y^{s}_{t,f}\|_2^2}{\|X^{s}_{t,f}\|_2^2} \\
        Loss = L_1({\rm OSSEM}(\hat{X^{s}}\cdot\alpha), Y^{s})
    \end{gather}
    where $X^{s}_{t,f}$ and $Y^{s}_{t,f}$ denote the noisy speech and the corresponding clean speech in the support set $s$. The ratio $\alpha$ is multiplied by the model output for only the loss computation. By adopting the feature re-scaling technique, the inner-loop gradients can be stabilized, and thus the SSM network can be tuned more accurately, finally enabling OSSEM to attain a better speaker adaptation performance.

\end{enumerate}

\section{Experimental Results}
\label{sec:ERA}
In this section, we investigate the effectiveness of OSSEM using the Voice Bank-DEMAND dataset. We evaluated the proposed OSSEM system using five standard metrics, PESQ, STOI, CSIG, CBAK, and COVL, which are commonly reported when using this dataset. We prepared 192-dimensional speaker embedding of one enrollment speech from each speaker using the ECAPA-TDNN. For fair comparisons, we trained all of our proposed systems using a total of 100 epochs (including pretrained epochs), with 1000 iterations in each epoch. In addition, we adopted the same set of hyperparameters of training. Note that we removed the enrolled speech from the two speakers in the testing set for a fair evaluation. OSSEM is a causal SE system that can perform enhancement in real time. 

\subsection{Speaker masks in the SE process}
\label{ssec:cmp1}
As shown in Fig. \ref{fig:BK diagram}, OSSEM uses the SSM network to directly modify the noisy input. To justify this early signal modification, we implemented four systems with the same SE and SSM networks, whereas the masks generated by the SSM network were applied in different places during the SE process: before the encoder, before the multi-head attention, before the decoder, and after the decoder. These systems are labeled Pre, Mid1, Mid2, and Last, respectively, and their PESQ scores on the test set are reported in Table \ref{tab:spk strat}. An addition system, called Non, represents a system having only an SE network and serves as a baseline for comparison. From Table \ref{tab:spk strat}, Pre performs better than the other systems, confirming that early signal modification is a suitable choice for building an adaptive SE system.
\vspace{-0.3cm}
\begin{table}[h]
\caption{Results of different places that the speaker masks (generated by the SSM network) are applied in the SE process.}
\centering
 \begin{tabular*}{6.3cm}{r|c c c c c}
 \hline
  & Pre & Mid1 & Mid2 & Last & Non \\ 
 \hline
 \hline
 PESQ & \textbf{2.80} & 2.78 & 2.76 & 2.64 & 2.69\\ 
 \hline
 \end{tabular*}
\label{tab:spk strat}
\vspace{-0.5cm}
\end{table}

\begin{table}[h!]
\caption{Results of OSSEM using different training techniques.}
\centering
 \begin{tabular*}{7cm}{l|c c} 
 \hline
 System & PESQ & STOI \\  
 \hline
 \hline
 Transformer$_{Spk}$ & 2.809 & 0.93 \\ 
 OSSEM with technique 1 & 2.818 & 0.93 \\
 OSSEM with techniques 1\&2 & 2.847 & 0.93 \\
 OSSEM with techniques 1\&2\&3 & 2.896 & 0.93 \\ 
 \hline
 \end{tabular*}
 \label{tab:ablation}
\vspace{-0.3cm}
\end{table}

\subsection{Ablation study}
\label{sec:cmp3}
Next, we investigate the effectiveness of the proposed training techniques introduced in Section \ref{ssec:OSSEMmethod}. Training techniques 1, 2, and 3 denote the speaker inner-loop, learning rate scaling, and feature re-scaling inner-loop, respectively. The transformer $_{Spk}$ in Table \ref{tab:ablation} denotes the SE network, which is first trained in a supervised manner and using Fig. \ref{fig:Training} (b) for online adaptation. From Table \ref{tab:ablation}, we can observe that each technique is helpful in improving the PESQ score. Among all of the proposed techniques, the feature re-scaling inner-loop (technique 3) is the most important.

\begin{table}[h]
\begin{threeparttable}
\caption{Evaluation results of OSSEM and other causal SE systems on the VoiceBank-DEMAND dataset.}
\centering
\small
\begin{tabular*}{\linewidth}{l|P{5mm} P{5mm} P{5mm} P{5mm} P{5mm} P{5mm}}
\hline
SE approach  & PESQ & CSIG & CBAK & COVL & STOI & causal\\
\hline
\hline
\textbf{Noisy\cite{pascual2017segan}} & 1.97 & 3.35 & 2.44 & 2.63 & 0.92 & --\\
\textbf{Wiener \cite{pascual2017segan}} & 2.22 & 3.23 & 2.68 & 2.67 & -- & yes\\
\textbf{Conv-TasNet \cite{koyama2020exploring}} & 2.53 & 3.95 & 3.08 & 3.23 & -- &  yes*\\
\textbf{Transformer \cite{fu2020boosting}}  & 2.69 & 4.07 & 3.03 & 3.38 & 0.93 & yes\\
\textbf{STFT-TCN \cite{koyama2020exploring}} & 2.73 & 4.11 & 3.25 & 3.42 & -- & yes*\\
\textbf{CRN-MSE \cite{tan2018convolutional}}  & 2.74 & 3.86 & 3.14 & 3.30 & 0.93 & yes\\
\textbf{TFSNN \cite{yuan2020time}}  & 2.79 & \textbf{4.17} & \textbf{3.27} & \textbf{3.49} & - & yes\\
\textbf{Transformer$_{Spk}$} & \textbf{2.81} & 3.95 & 3.19 & 3.36 & 0.93 & yes \\
\hline
\hline
\textbf{OSSEM} & \textbf{2.90} & \textbf{4.11} & \textbf{3.15} & \textbf{3.50} & \textbf{0.93} & yes\\
\hline
\end{tabular*}
\footnotesize{yes* denotes the use of look-ahead mechanism, where \textbf{Conv-TasNet} uses 1ms look-ahead, and \textbf{STFT-TCN} uses 4ms look-ahead.}
\label{tab:denoise}
\end{threeparttable}
\end{table}

\subsection{Comparison of OSSEM with other causal SE systems}
\label{ssec:cmp2}
Table \ref{tab:denoise} lists the results of OSSEM and several related SE systems, including the modified Transformer SE network \cite{fu2020boosting}, Transformer$_{Spk}$ (as discussed in Section \ref{sec:cmp3}), and some well-known causal SE systems. Note that our testing set is a slightly modified version of the original testing set of Voice Bank-DEMAND. In our preliminary experiments, we have confirmed that the results of Noisy, Wiener, and Transformer tested on the modified testing sets are almost identical to those tested on the original Voice Bank-DEMAND. Therefore, we directly copied the results of Noisy, Wiener, and other SE systems from existing studies listed in Table \ref{tab:denoise} for comparison.
\par
From Table \ref{tab:denoise}, we first note that OSSEM outperforms both Transformer and Transformer$_{Spk}$, confirming the effectiveness of one-shot adaptation through meta-learning. Next, it was observed that OSSEM outperforms other systems in most of the five metric scores, confirming the effectiveness of OSSEM as compared to other systems. We also noticed that, among the five metrics, OSSEM yields a slightly lower yet comparable performance to DEMUCS \cite{defossez2020real}. We argue that the model size of DEMUCS is almost twice that of the OSSEM system (OSSEM: 38MB; DEMUCS: 73MB \cite{lee2021demucs}). Thus, OSSEM achieves a better applicability to real-world systems, particularly when storage is limited. We also note that DEMUCS uses a 3-ms look-ahead mechanism, whereas our OSSEM system is completely causal. To further investigate the capabilities of OSSEM, we implemented a non-causal version of OSSEM. The results of the non-causal OSSEM do not notably outperform the causal version, probably owning to the fact that speaker embedding can mitigate the information gap between causal and non-causal systems.

\subsection{Analysis of the SSM network}
\label{spkana}
\begin{figure}[h]
    \centering
    \includegraphics[width=8cm]{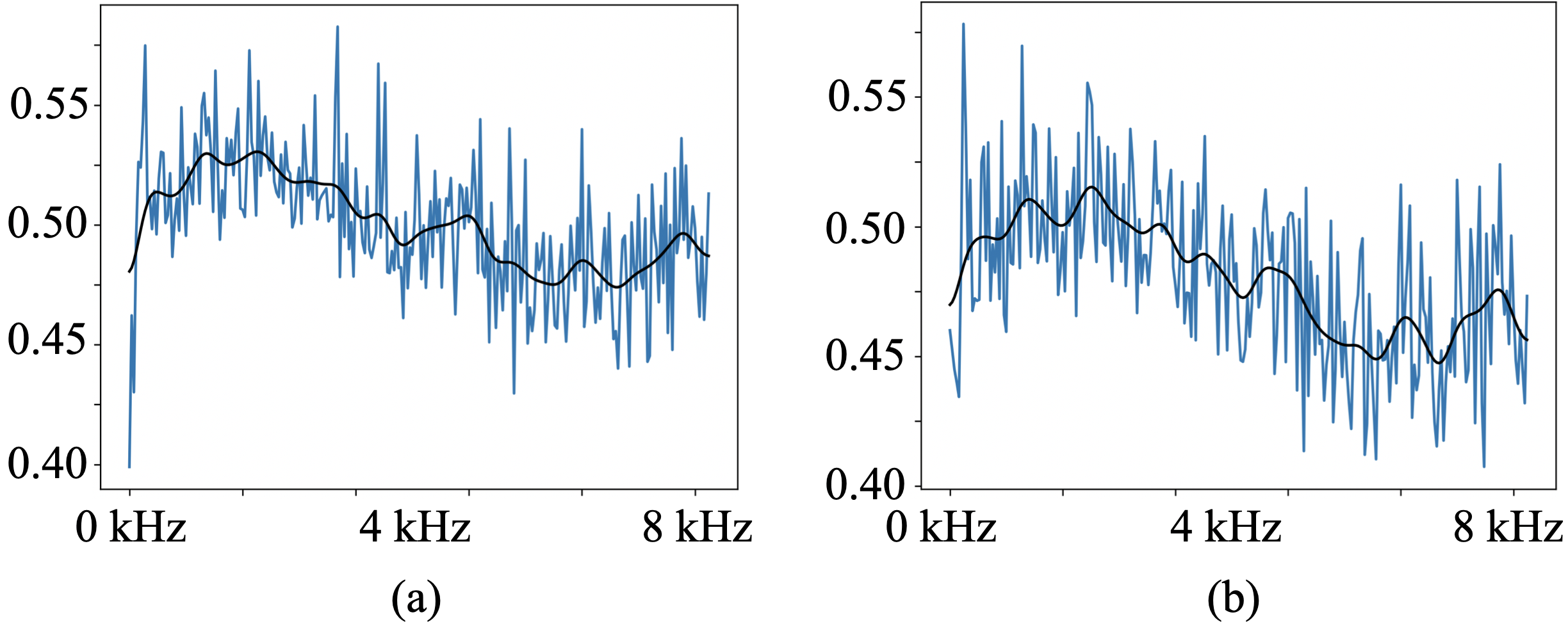}
    \vspace{-0.3cm}
    \caption{The masks generated by the SSM network of speech samples from (a) a female speaker, and (b) a male speaker.}
    \label{fig:spkana}
\end{figure}

In this section, we further investigate the function of the SSM network in OSSEM. Figs. \ref{fig:spkana} (a) and (b) show the mask plots (output of the SSM network) of female and male test speakers, where the x- and y- axes of the plots indicate the frequency and amplitude, respectively. The blue and black lines indicate the detailed mask values for every frequency bin and the smoothed mask value over the frequency (showing the trend of the mask). From Fig. \ref{fig:spkana}, we first note that the frequency range of the first peak in Fig. \ref{fig:spkana} (a) is higher than that of Fig. \ref{fig:spkana} (b), confirming that the mask indeed captures the gender information. Next, we observed that Figs. \ref{fig:spkana} (a) and (b) present extremely different patterns, showing that for each speaker, a specified mask is generated and assigned, to modify the noisy input.

\begin{figure}[h]
    \centering
    \includegraphics[width=7.6cm]{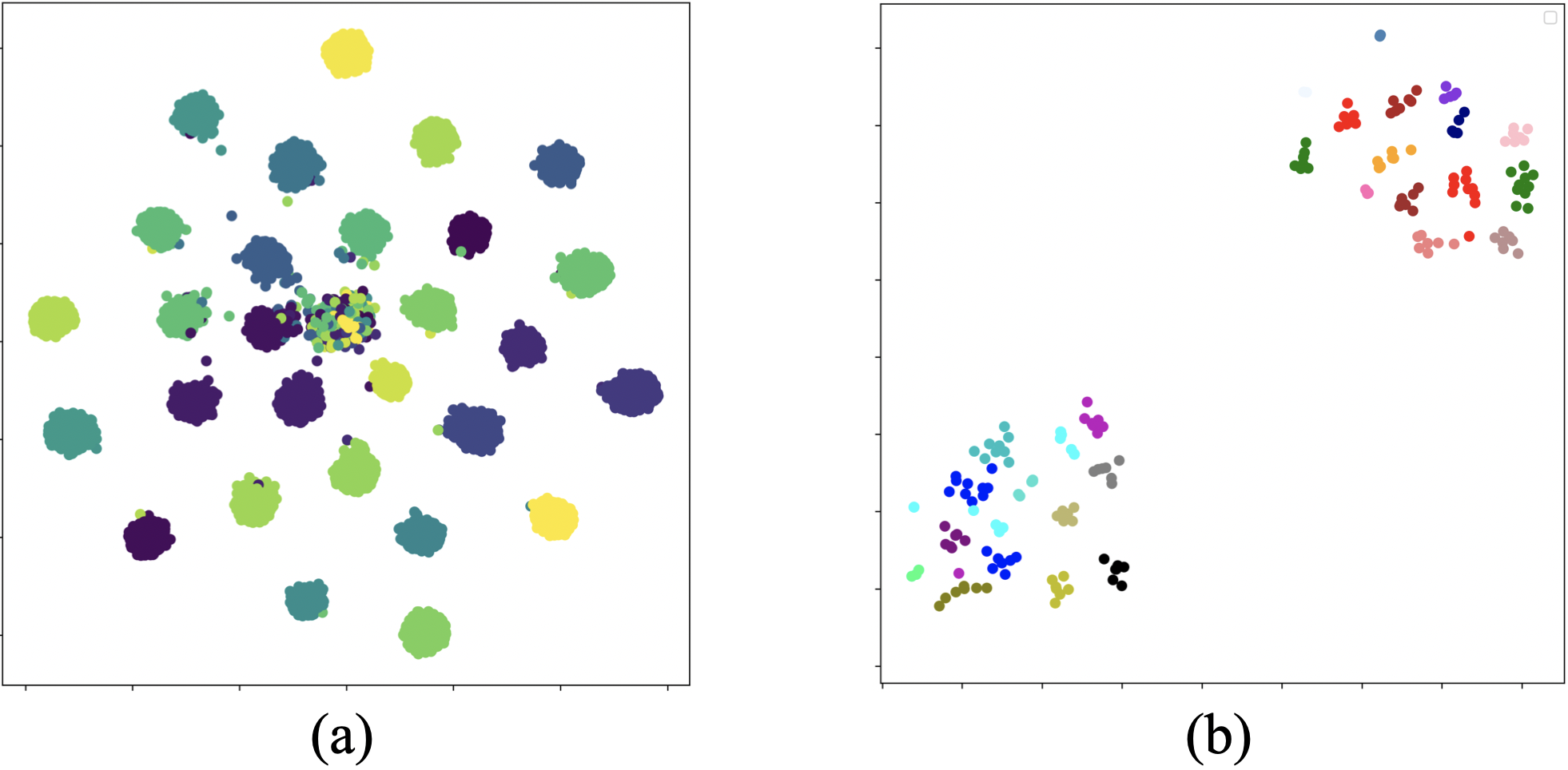}
    \vspace{-0.3cm}
    \caption{The t-SNE analysis of (a) ECAPA-TDNN speaker embedding of all speakers, and (b) the speaker masks of all speakers.}
    \label{fig:embana}
\vspace{-0.3cm}
\end{figure}

Fig. \ref{fig:embana} (a) shows the results of a t-SNE analysis \cite{tsne} on the embedding vectors extracted using ECAPA. The speech utterances were taken from the training set of Voice Bank-DEMAND, which comprises utterances from 28 speakers. From Fig. \ref{fig:embana} (a), we can note that the utterances from each speaker are grouped together, and there are a total of 28 groups. Moreover, we can observe a clear distance between each pair of groups. The results confirm that the speaker representations extracted by ECAPA are informative and discriminative. Fig. \ref{fig:embana} (b) shows the results of a t-SNE analysis that was performed on the masks generated by the SSM network with the same utterances used in Fig. \ref{fig:embana} (a). Note that the masks are also similarly informative and discriminative but are divided into two groups of two genders. The transition from Fig. \ref{fig:embana} (a) to Fig. \ref{fig:embana} (b) suggests that the SSM network further emphasized the characteristics of gender when converting the embedding into masks. The results conform to a previous study \cite{kolbaek2016speech} showing that the gender identity of the speaker plays a major factor in SE performance.    

\section{Conclusion}
\label{sec:conclude}
In this study, we proposed an OSSEM system that can adapt a pretrained SE model to a new speaker in a one-shot manner. Experiment results confirm the effectiveness of speaker-specific masking and metal learning for fast SE model adaptation. The results also show that OSSEM outperforms several well-known causal SE systems in terms of the standard evaluation metrics. In summary, the contributions of this study are threefold: (1) The proposed OSSEM is the first one-shot speaker adaptive SE system based on meta-learning, (2) a novel SSM network that directly modifies the noisy input is proposed and was proven to be effective, and (3) several techniques to improve the stability of meta-learning have been adopted and modified to conform to the SE task. In the future, we will explore the application of the proposed approach to other speech generation tasks, such as speech separation and target speaker extraction.

\vfill\pagebreak


\bibliographystyle{IEEEbib}
\begin{spacing}{0.8}
\bibliography{refs}
\end{spacing}
\end{document}